\documentclass[preprint,prd,nofootinbib,tightenlines,amsmath]{revtex4}
\usepackage{graphicx}% Include figure files
\usepackage{bm}% bold math
\usepackage{slashbox}

\begin{document}
\baselineskip=15pt \parskip=5pt

\vspace*{3em}

\title{Gas Model of Gravitons with Light Speed }

\author{ Ming Chen $^{1,}$}
\email{mchen1905@hotmail.com}
\author{Yong-Chang Huang$^{1,}$$^{2,}$}
\email{ychuang@bjut.edu.cn}
\affiliation{$^1$Institute of Theoretical Physics, Beijing University of Technology, Beijing 100124, China\\
$^2$CCAST (World Lab.), P.O. Box 8730, Beijing 100080, China}

\date{\today $\vphantom{\bigg|_{\bigg|}^|}$}

\begin{abstract}
\vspace{1em}
We first review some aspects of gravitational wave and the thermodynamic expression of Einstein field equations, these achieved conclusions allow people to think of Einstein's gravitational wave as a kind of sound wave in ordinary gas which propagates as an adiabatic compression wave. In the following, using the properties of photon gas in ``white wall box'', we find an analogous relationship between ordinary gas and photon gas through sound velocity formula. At last, by taking the ordinary gas as an intermediary, we find that gravitational wave is analogous to photon gas, or equally, gravitons are analogous to photons although they are different in some ways such as spins and coupling strengths, and these different properties don't affect their propagation speeds. Utilizing this analogous relationship, we achieve the gas model of gravitons and this model naturally gives out the light speed of gravitons . \\[2cm]



\end{abstract}

\maketitle

\section{Introduction}
It is well known that, Einstein's field equation is a milestone of theoretical physics. It gives the universe a geometric background which overturns our traditional notion so much. It is no exaggeration to say that General Relativity sets a new tone for what a physical theory can be, and has truly revolutionized our understanding of the universe \cite{1} \cite{2}. We have also known there are a great many similarities between gravitation and electromagnetism. It should therefore come as no surprise that Einstein's equations, like Maxwell's equations, have radiative solutions. The theory of gravitational radiation is complicated by the nonlinearity of Einstein's equations. And any gravitational wave is itself a distribution of energy and momentum that contributes to the gravitational field of the wave. This complication prevents our being able to find general radiative solutions of the exact Einstein equations.

There are two approaches to this difficulty. One is to study only the weak-field radiative solutions of Einstein equations, which describe waves carrying not enough energy and momentum to affect their own propagation. The other approach is to look long and hard for special solutions of the exact Einstein field equations. A great deal of mathematical ingenuity has gone into the second approach, with results of some elegance. Gravitational waves are more complicated than electromagnetic waves because they contribute to their own source outside the material gravitational antenna. However, the simple properties of both electromagnetic and gravitational waves emerge when we look far out into the wave zone, where the fields are weak. This is the right reason that we choose the first approach here.

This means that, according to the approximated linear expressions of Einstein field equations, it is natural to consider the corresponding relationship between classical propagating waves (like the radiative solutions of Maxwell's equations) and gravitational waves. Many people try to utilize this relationship to understand the profound gravitational field. This gives out the implication of the analogy of the gravitational waves and classic propagating waves.

The arrangements of the paper are: In Section II, we introduce several ways of understanding the gravitational wave. Then we find the analogous relationship between it and the adiabatic compression wave. In Section III we introduce the photon gas in the ``white wall box''. By use of the properties of it, we build an equivalence between ordinary gas and photon gas through the sound velocity formula . Then we make a generalization of Section II to get the analogy between photon gas and gravitational wave, arriving at the gas model. Section IV gives some comments, conclusions and outlooks.

\section{Gravitational waves}
We have known that one of the most fundamental concepts in Special Relativity is that the space-time interval $dS$ between any two neighboring points is given by the expression
\begin{equation}
ds^{2}=\eta_{\mu\nu}dx^\mu dx^\nu.
\end{equation}

The same physical concept is carried over into General Relativity, with one key difference: space-time is no longer necessarily the ``flat" space-time described by the Minkowski metric, but will in general be curved in order to represent what we call gravitation. The more general statement of the definition of the space-time interval is
\begin{equation}
ds^{2}=g_{\mu\nu}dx^\mu dx^\nu.
\end{equation}

Since all of the information about space-time curvature is encoded in the metric $g_{\mu\nu}$. We only deal with a small perturbation to flat space-time. Then it makes sense to write the metric in the form
\begin{equation}
g_{\mu\nu}=\eta_{\mu\nu}+h_{\mu\nu},
\end{equation}
where $h_{\mu\nu}$ represents the metric perturbation away from Minkowski space.

The key physics of the problem is thus carried in the form of $h_{\mu\nu}$. It is a remarkable fact that in the weak-field limit, the non-linear Einstein equations can be approximated as linear equations. There are a great deal of gauge freedoms in the construction of explicit forms for $h_{\mu\nu}$, and they are the source of some confusions. To make the decomposition of the metric into a flat background plus a perturbation be unique, we have to add some necessary gauge, the so-called transverse traceless gauge, or ``TT gauge" for short. In this gauge, coordinates can be marked out by the world lines of freely-falling test masses.

Thus, the equation of motion is then
\begin{equation}
\partial^\mu \partial^\nu h^{TT}_{\mu\nu}=0.
\end{equation}

From the wave Eq.(4) we begin finding solutions. Those familiar with the analogous problem in electromagnetism will notice that the procedure is almost precisely the same. A particularly useful set of solutions to this wave equation are the plane waves, given by
\begin{equation}
h_{\mu\nu}^{TT}=C_{\mu\nu}e^{ik_{\sigma}x^{\sigma}},
\end{equation}
where $C_{\mu\nu}$ is a constant, symmetric, $(0,2)$ tensor, which is obviously traceless and purely spatial
\begin{equation}
C_{0\nu}=0,
\end{equation}
\begin{equation}
\eta^{\mu\nu} C_{\mu\nu}=0.
\end{equation}

The plane wave Eq.(5) is therefore a solution to the linearized equation if the wave vector is null, this is loosely translated into the statement that gravitational waves propagate at the speed of light. Of course, there are the linear plane waves corresponding to some exact solutions of the nonlinear equations that haave similar properties, these solutions are a class called plane-fronted waves with parallel rays, or PP waves for short \cite{3}.

Clearly, what we talk about above cares only about the geometrical background which leads the plane wave solution to the geometrical meaning. Ted Jacobson gives out the different understanding about gravitational wave \cite{4}. He doesn't care about the geometrical background, so the system is based on any causal horizon. The fundamental principle at play in his analysis is: the equilibrium thermodynamic relation $\delta Q=TdS$, as interpreted in terms of energy flux and area of local Rindler horizons, can be satisfied only if gravitational lensing by matter energy distorts the causal structure of spacetime in just such a way that the Einstein equation holds.

The local equilibrium conditions used in Ref.[4] are somehow equivalent to the weak-field limit. So we have, in the Einstein equation, a system of local partial differential equations that are time reversal invariant and whose solutions include propagating waves. In other words, the new insight of Einstein equation gives the thermodynamic property to the wave solution. Moreover, the thermodynamic deduction of the Einstein equation of state presumes the existence of local equilibrium conditions under that the relation $\delta Q=TdS$ applies only to variations between nearby states of local thermodynamic equilibrium. This allows us to think of the propagating wave as analogous to sound in a gas, which propagats as an adiabatic compression wave no matter what kind of variations they might be, because what we care about here is the energy transferred \cite{5}.

In other words, there is an analogous relationship between gravitational wave and sound wave. Meanwhile we have learned to describe the fundamental observables of microscopic phenomena in terms of elementary particles and their collisions. In classic electrodynamics it is the plane wave solutions of Maxwell's equations that lead most naturally to an interpretation in terms of a particle, the photon. Similarly, it is the radiation solutions of Einstein's equations that will lead here to the concept of a particle of gravitational radiation, the graviton \cite{6}. So this analogy is equally an analogy between gravitons and the gas molecules. This is consistent with our argument in Section I.

Though we know a little about the gravitons, we can still realize that this analogy makes little sense. We cannot only determine the velocity of gravitons but also any meaningful properties of them. As a result of this, we have to find another way to solve this trouble.

\section{Photon gas in the ``white wall box''}
Following Sec.II, we want to find some other things better than gas molecule to simulate gravitons. In this section we will get help from photon gas by use of sound velocity formula.

According to Ref.\cite{7} \cite{8}, one can get a $2-dimensional$ ``white wall box'' for photon gas, in this system, thermalization is achieved in a photon-number-conserving way by photon-scattering-off dye molecules, and the cavity mirrors provide both an effective photon mass and a confining potential.

The Bose-Einstein distribution factor for this type of photon gas is
\begin{equation}
f(\mu,T,\varepsilon)=\frac{1}{exp\left(\frac{\varepsilon-\mu}{k_{B}T}\right)-1}
\end{equation}
where the transversal energy, or the effective energy is $\varepsilon=\varepsilon^\prime-\hbar \omega_{cutoff}=\hbar \omega$,  $\mu$ is the chemical potential.

In the $2-dimensional$ ``box'', the state numbers between energy $\varepsilon \rightarrow \varepsilon + d\varepsilon$ is
\begin{equation}
dn=\frac{A}{h^{2}}2\pi p*dp*=\frac{2\pi A}{c^{2}h^{2}}\varepsilon d\varepsilon
\end{equation}
where the symbol $A$ is the area of the ``box'' . So, the photon numbers is,
\begin{equation}
dN(\varepsilon)=\frac{4\pi A\varepsilon d\varepsilon}{c^{2}h^{2}exp\left(\frac{\varepsilon-\mu}{k_{B}T}\right)-1}
\end{equation}
and the energy is,
\begin{equation}
U(\varepsilon,\mu,T)d\varepsilon=\varepsilon dN(\varepsilon)=\frac{4\pi A\varepsilon^{2}d\varepsilon}{c^{2}h^{2}exp\left(\frac{\varepsilon-\mu}{k_{B}T}\right)-1}
\end{equation}
thus the total energy of all the frequencies is,
\begin{equation}
U(\varepsilon,\mu,T)=\frac{4\pi A}{c^{2}h^{2}}\int_{0}^{\infty}\frac{\varepsilon^2 d\varepsilon}{exp\left(\frac{\varepsilon-\mu}{k_{B}T}\right)-1}
\end{equation}
using the following equation,
\begin{equation}
\frac{1}{exp(x)-1}=\overset{\infty}{\underset{j=1}{\sum}}exp(-jx)
\end{equation}
we can get
\begin{equation}
U(\mu,T)=\frac{8\pi Ak_{B}^{3}T^{3}}{c^{2}h^{2}}\overset{\infty}{\underset{j=1}{\sum}}exp\left(\frac{j\mu}{k_{B}T}\right)\frac{1}{j^{3}}
\end{equation}

If $exp\left(\frac{j\mu}{k_{B}T}\right)$ is very small, we can make the suitable approximation by keeping the first two orders of the expansion.
\begin{equation}
U(\mu,T)\cong\frac{8\pi Ak_{B}^{3}T^{3}}{c^{2}h^{2}}\overset{\infty}{\underset{j=1}{\sum}}\left(\frac{1}{j^{3}}+\frac{1}{j^{2}}\frac{\mu}{k_{B}T}\right)=\frac{2.404k_{B}^{3}}{\pi c^{2}\hbar^{2}}AT^{3}+\frac{\pi k_{B}^{2}}{3c^{2}\hbar^{2}}\mu AT^{2}
\end{equation}

According to Ref.\cite{7}, when $\mu \rightarrow 0$ £¬and $T$ is the limited value, we can
neglect the term including the factor $\mu$ from the viewpoint of calculation, so we get
\begin{equation}
U(\mu,T)\cong\bar{\sigma}AT^{3}
\end{equation}
where $\bar{\sigma}=\frac{2.404k_{B}^{3}}{\pi c^{2}\hbar^{2}}$  is the Stefan-Boltzmann constant in $2-dimensional$ space.

Therefore, we can get the pressure and the entropy as follows,
\begin{equation}
p={\frac{1}{2}\bar\sigma}T^{3}
\end{equation}
\begin{equation}
S=\frac{3}{2}\bar{\sigma}AT^{2}
\end{equation}

Rrewritting Eq.(18) as $S=3\left(\frac{\bar{\sigma}}{2}\right)^{\frac{1}{3}}P^{\frac{2}{3}}A$, then,
\begin{equation}
dS=3\left(\frac{\bar{\sigma}}{2}\right)^{\frac{1}{3}}\left(\frac{2}{3}P^{-\frac{1}{3}}AdP+P^{\frac{2}{3}}dA\right)
\end{equation}
thus, the adiabatic compression coefficient is,
\begin{equation}
\kappa_{s}=-\frac{1}{A}\left(\frac{\partial A}{\partial P}\right)_{s}=\frac{2}{3P}
\end{equation}

We know that two versions of the sound velocity formula are,
\begin{eqnarray}
v=\sqrt{\left(\frac{\partial P}{\partial\rho}\right)_{s}}=\sqrt{\gamma\frac{P}{\rho}}
\end{eqnarray}
\begin{equation}
v=\sqrt{\frac{1}{\rho\kappa_{s}}}
\end{equation}
where $\rho=\frac{m}{V}$ is the density, $m$ is the mass, $V$ is the volume, and $P$ is the pressure of the medium. The subscript $_s$ indicates the adiabatic progress and $\gamma=\frac {C_{P,m}}{C_{V,m}}$ is the ratio of specific heat capacity (See appendix A for details).

If we put Eq.(20) into Eq.(22), we get,
\begin{equation}
v=\sqrt{\frac{3}{2}\frac{P}{\rho}}
\end{equation}

We now consider the photon gas in $n-dimensional$ space, the relationship between the angular frequency $\omega$ and the quantum state numbers is,
\begin{equation}
D(\omega)\propto\omega^{n-1}
\end{equation}

So the total energy of $n-dimensional$ photon gas is,
\begin{equation}
U=\intop_{0}^{\infty}\varepsilon f(\omega,T)D(\omega)d\omega\propto T^{n+1}
\end{equation}
we also know, from the Stefan-Boltzmann law, that the relationship between the temperature and the energy of the photon gas is, $U\propto T^{\gamma}$.
Comparing Eq.(25) and this expression, we can get, $\gamma=n+1.$
This reminds us that Eq.(23) is the special form in $2-dimensional$ space, the general formula should be,
\begin{equation}
v=\sqrt{\frac{\gamma}{n}\frac{P}{\rho}}
\end{equation}

If we define, $c=\sqrt{n}v$,
then,
\begin{equation}
c=\sqrt{\gamma\frac{P}{\rho}}
\end{equation}

Comparing Eq.(27) with Eq.(21), we will find that when we get Eq.(21), we use the equation of adiabatic progress, i.e., the Poisson formula $PV^\gamma=const.$ (Given by Eq.(29)). Here the fact that photon gas satisfies the equation of sound velocity tries to remind us that the photon gas is somehow a kind of sound gas. Under compression, it can propagate in the form of adiabatic compression just like the sound.

Furthermore, from Section II, we have known that the universal nature of gravity is also demonstrated by the fact that its basic equations closely resemble the laws of thermodynamics and hydrodynamics. Complying with this, it's legal to generalize the analogy between gravitational wave and adiabatic compression wave to the analogy between photon gas and gravitational wave. This generalization consummates the gas model. As we can take this analogy equally as the analogy between photons and gravitons, this means that the gas model gives some properties of photons to gravitons.

The relationship between photon and graviton has already been investigated in Ref.\cite{9}. They view the general relativity as a quantum theory which propagates a unique kind of weakly coupled quantum particles with zero mass and $spin-2$, and assume black holes can achieve Bose-Einstein condensation, then expect that some of its properties have counterparts in ordinary Bose-Einstein condensation such as in system of photons. Compared, we begin, in the different way, from the wave side. We first make clear that the gravitational wave is analogous to the classic propagating wave. Then we prove that the photon gas is also like the classic propagating wave, so we can relate the gravitational wave to the photon gas. This is the essential difference between our work and theirs.

In fact, when the fields are strong, we can also get the plane wave solution of the gravitational field. We know that the Rindler transformations work for a wide variety of spherically symmetric solutions to gravitational field equations. We can see that, the condition $-g_{00}=N^2\rightarrow 0$ on the horizon can give out the plane wave solutions of any classical (or quantum) field equations near the horizon \cite{10}. This agrees with the plane wave solution of the gravitational field in the weak-field limit. But we don't know how photons act in the strong field, so we cannot generalize the gas model from weak field to strong field directly.

\section{Some comments, conclusions and outlooks}

In summary, we generalize the analogy between gravitational wave and adiabatic compression wave to the analogy between photon gas and gravitational wave, or equally, the analogy between photons and gravitons. This analogy is referred as the gas model of gravitons.

Many physicists believe that gravity and space-time geometry are emergent. Also string theory and its related developments have given several indications in this direction. Particularly important clues come from the Ads/CFT, or more generally, the open/closed string correspondence. This correspondence leads to a duality between theories that contain gravity and those that don't. It therefore provides evidence for the fact that gravity can emerge from a microscopic description that doesn't know about its existence \cite{11}. Complying with these ideas, we here point out one more possibility to interpret gravity.

Apparently, photon is different from graviton in many ways. They are particles transferred between different interactions and have different coupling constants. Meanwhile, we know that the allowed spin measurement values represent the actually measured degrees of freedom. And both photons and gravitons are massless particles with spins 1 and 2 respectively. The $spin-2$ gravitons have two separative measured values, i.e., -2, 2. Similarly, the $spin-1$ photons also have two separative measured values i.e., -1, 1. So gravitons and photons have the same degrees of freedom. This strengthens this model in the statistical side.

The easy check of this model is the speed of gravitational wave or gravitons. Under this generalization, gravitons are analogous to photons from the viewpoint of statistics, which is quite natural because what we care about is the energy in the first place. So the speed of graviton is just the speed of light.
Furthermore, we know that Hawking radiation and Bekenstein-Hawking entropy are the two robust predictions of a yet unknown quantum theory of gravity. Any theory which fails to reproduce these predictions is certainly incorrect. This will guide us in the future deductions and we want to check this model with the entropy of black hole. We will continue the comments by adding some properties of photons to gravitons. And these works will be done in next paper.

\begin{center}
\large Appendix
\end{center}

\subsection{Sound velocity}
The sound wave in the gas is propagated by adiabatic compression and expansion periodically of the local gas medium. The velocity of the sound wave depends on the density and the elastic property of the medium. The sound velocity formula is
\begin{equation}
\nu=\sqrt{\left(\frac{\partial P}{\partial\rho}\right)_{s}},
\end{equation}
where $\rho=\frac{m}{V}$ is the density£¬$m$ is the mass,$V$ is the volume,and $P$ is the pressure of the medium. The subscript  $_s$  indicates the adiabatic progress.

Meanwhile we also know the equation of adiabatic progress, i.e, the Poisson formula
\begin{equation}
PV^\gamma=const. ,
\end{equation}
where $\gamma=\frac {C_{P,m}}{C_{V,m}}$ is the ratio of specific heat capacity.We can rewrite Eq.(29) to be
\begin{equation}
P=const.\rho^\gamma .
\end{equation}

Then we have
\begin{eqnarray}
\nu=\sqrt{\left(\frac{\partial P}{\partial\rho}\right)_{s}}=\sqrt{\gamma \frac{P}{\rho}}.
\end{eqnarray}

From the other side, we know
\begin{eqnarray}
v^{2}=(\frac{\partial P}{\partial\rho})_{s}=(\frac{\partial P}{\partial T})_{s}(\frac{\partial T}{\partial V})_{s}\frac{dV}{d\rho}=\frac{(\frac{\partial S}{\partial T})_{P}(\frac{\partial S}{\partial V})_{T}}{(\frac{\partial S}{\partial P})_{T}(\frac{\partial S}{\partial T})_{V}}\frac{dV}{d\rho}=\frac{(\frac{\partial S}{\partial T})_{P}}{(\frac{\partial S}{\partial T})_{V}}(\frac{\partial P}{\partial V})_{T}\frac{dV}{d\rho}=\frac{C_{P}}{C_{V}}\frac{V}{\kappa_{T}m}=\frac{1}{\rho\kappa_{s}},
\end{eqnarray}
this means
\begin{equation}
\nu=\sqrt{\frac{1}{\rho \kappa_s}} .
\end{equation}
where the symbol $\kappa_s$ is the adiabatic compression coefficient.

\vspace{3em}
\begin{center}
\large Acknowledgements
\end{center}
This work is supported by National Natural Science Foundation of China (No.11275017 and No.11173028).

\end{document}